\documentclass{aa}
\usepackage{graphicx}
\usepackage{natbib} \bibpunct{(}{)}{;}{a}{}{,}
\usepackage{color}
\usepackage{txfonts}

   \def\HeI{\ion{He}{i}}
 \def\HeII{\ion{He}{ii}} \def\HeIII{\ion{He}{iii}} \def\CI{\ion{C}{i}}
 \def\CII{\ion{C}{ii}} \def\CIII{\ion{C}{iii}} \def\CIV{\ion{C}{iv}}
 \def\CV{\ion{C}{v}}  
   
  \def\OI{\ion{O}{i}} \def\OII{\ion{O}{ii}}
 \def\OIII{\ion{O}{iii}} \def\OIV{\ion{O}{iv}} \def\OV{\ion{O}{v}}
 \def\OVI{\ion{O}{vi}} \def\OVII{\ion{O}{vii}} 
 \def\SiIII{\ion{Si}{iii}} \def\SiIV{\ion{Si}{iv}} \def\SiV{\ion{Si}{v}}
  \def\FeII{\ion{Fe}{ii}} \def\FeIII{\ion{Fe}{iii}}
 \def\FeIV{\ion{Fe}{iv}} \def\FeV{\ion{Fe}{v}} \def\FeVI{\ion{Fe}{vi}}
 \def\FeVII{\ion{Fe}{vii}} \def\FeVIII{\ion{Fe}{viii}} \def\FeIX{\ion{Fe}{ix}}
 \def\FeX{\ion{Fe}{x}} \def\FeXI{\ion{Fe}{xi}} \def\FeXII{\ion{Fe}{xii}}
 \def\FeXIII{\ion{Fe}{xiii}} \def\FeXIV{\ion{Fe}{xiv}} \def\FeXV{\ion{Fe}{xv}}
 \def\FeXVI{\ion{Fe}{xvi}} \def\FeXVII{\ion{Fe}{xvii}}
\def\dif{\mbox{d}} \newcommand{\msunpyr}{\,M_\odot\,\mbox{yr}^{-1}}
\newcommand{\dint}{\,\mbox{d}} 
\newcommand{\ddif}{\mbox{d}}
\newcommand{\kms}{\ifmmode{\,\mbox{km}\,\mbox{s}^{-1}}\else{km/s}\fi}
\newcommand{\msun}{\ifmmode M_{\odot} \else M$_{\odot}$\fi}
\newcommand{\rsun}{\ifmmode R_{\odot} \else R$_{\odot}$\fi}
\newcommand{\lsun}{\ifmmode L_{\odot} \else L$_{\odot}$\fi}
\newcommand{\zsun}{\ifmmode Z_{\odot} \else $Z_{\odot}$\fi}
\newcommand{\velo}{\ifmmode\varv\else$\varv$\fi}
\newcommand{\vinf}{\ifmmode\velo_\infty\else$\velo_\infty$\fi}
\begin{document} 
 
\title{Hydrodynamic model atmospheres for WR\,stars:} \subtitle{Self-consistent
  modeling of a WC star wind}
  
\titlerunning{Hydrodynamic model atmospheres for WR\,stars}
 
\author{G.\ Gr\"{a}fener\inst{1,2,3} \and W.-R.\ Hamann\inst{1}}
 
\institute{ Institut f\"ur Physik, Astrophysik, Universit\"at Potsdam, Am Neuen
  Palais 10, D-14469 Potsdam, Germany \and Institute of Astronomy, ETH Zentrum
  SEC , Scheuchzer Str. 7, CH-8092 Z\"urich, Switzerland \and PMOD/WRC, CH-7260
  Davos Dorf, Switzerland}

\offprints{G.\ Gr\"afener\\
  \email{goetz@astro.physik.uni-potsdam.de}}
 
\date{Received ; Accepted}
 
\abstract{We present the first non-LTE atmosphere models for WR\,stars that
  incorporate a self-consistent solution of the hydrodynamic equations.  The
  models account for iron-group line-blanketing and clumping, and compute the
  hydrodynamic structure of a radiatively driven wind consistently with the
  non-LTE radiation transport in the co-moving frame. We construct a
  self-consistent wind model that reproduces all observed properties of an
  early-type WC\,star (WC5).  We find that the WR-type mass-loss is initiated at
  high optical depth by the so-called `Hot Iron Bump' opacities ({\FeIX}--{\sc
    xvi}).  The acceleration of the outer wind regions is performed by
  iron-group ions of lower excitation in combination with C and O.
  Consequently, the wind structure shows two acceleration regions, one close to
  the hydrostatic wind base in the optically thick part of the atmosphere, and
  another farther out in the wind. In addition to the radiative acceleration,
  the `Iron Bump' opacities are responsible for an intense heating of deep
  atmospheric layers.  We find that the observed narrow {\OVI}-emissions in the
  optical spectra of WC\,stars originate from this region.  By their dependence
  on the clumping factor we gain important information about the location where
  the density inhomogeneities in WR-winds start to develop.
  
   }

\maketitle

\section{Introduction} 
\label{sec:intro}

The modeling of the atmospheres of Wolf-Rayet (WR) stars with their strong
stellar winds underwent a strong development during the last two decades.
Starting with pure-helium atmospheres \citep{hil1:87,hil2:87,ham2:87}, the main
goal was to augment the models by more complex atoms like C, N, and O
\citep[e.g.][]{hil1:89,koe1:95}, until a major step towards a realistic
determination of stellar parameters and emergent flux distributions was
performed by the inclusion of iron-group line-blanketing
\citep{hil1:98,hil1:99,gra1:02} and clumping \citep[see][]{ham1:98}.

For the inclusion of line-blanketing major revisions of the model codes were
necessary. In line-blanketed models the whole spectral range is covered by
millions of overlapping lines. Due to the frequency coupling in expanding
atmospheres, no sampling or re-ordering techniques can be applied \citep[cf.\ 
the technique of opacity distribution functions, e.g.][]{car1:79}, i.e., the
complete radiation field must be calculated on a fine frequency grid, with each
line transition at its proper place.

As a by-product of these numerically expensive calculations also the radiative
acceleration on the wind material is obtained. This is of special interest for
WR\,stars, because it is not clear whether their strong winds can be driven by
radiation pressure alone. In contrast to the thin winds of OB\,stars, WR\,winds
show wind momenta $\dot{M}\vinf$ far above the single scattering limit $L/c$.
This means that, on average, each photon leaving the star must be scattered
several times to accelerate the wind to the observed terminal velocities.

The first results from line-blanketed atmosphere calculations have shown that
the acceleration of the outer part of a WR\,wind is easily explained by
radiative driving, whereas the models failed completely in the deeper layers
where the mass-loss is initiated \citep{gra1:00,her1:01,gra1:02,hil1:03}. On the
other hand, \citet{nug1:02} and \citet{lam1:02} demonstrated by a critical-point
analysis that the {\em mass-loss rates} of WR\,stars can probably be maintained,
if the critical point of the equation of motion is located in the optically
thick part of the atmosphere at a temperature close to 160\,kK.  In this case
the mass-loss is initiated by the so-called `Hot Iron Bump' -- an increase of
the Rosseland mean opacity, which is mainly due to the excitation of the Fe
M-shell ions ({\FeIX}--{\sc xvi}).  Due to their method of analysis, however, it
remained unclear whether the assumed conditions are actually met in
WR\,atmospheres, and if the wind can be further accelerated beyond the optically
thick regime.

Due to their high wind densities and their hard radiation field, WR\,stars
develop extended atmospheres in extreme non-LTE that require sophisticated
modeling techniques.  Their complex ionization structure is strongly affected by
the interplay between different elements \citep[see][]{gra1:02}. It is therefore
necessary to employ a full non-LTE treatment and a radiation transport that
correctly accounts for line-overlaps.  Moreover, WR\,winds show high velocities
at large optical depths, i.e., complete atmosphere models including an accurate
solution of the temperature structure are required.  The numerical cost of our
present models is therefore much higher than for already existent stellar wind
models \citep[e.g.\ 
][]{cas1:75,abb1:85,pau1:86,luc1:93,sch1:94,pau1:94,vin1:99,pau1:01,krt1:04}
which are mostly focusing on the thin winds of OB\,stars.

An additional important aspect for the driving of WR\,winds, which is not
included in any of the previous wind models, is the role of density
inhomogeneities.  For WR\,stars, these are directly observed as time-dependent
variations of the line profiles \citep{lep1:99,lep2:99,lep1:00,koe2:01}.  They
are included in our models in the form of a clumping factor $D$, which lowers
the empirically derived mass-loss rates by a factor $\sqrt{D}$
\citep[see][]{ham1:98}, i.e., it reduces the wind momentum problem for
WR\,stars.  Typically, values in the range of $D$\,=\,4--10 are obtained by a
fit of the electron scattering wings of strong emission lines.  An effect of
similar importance is the influence of clumping on the {\em radiative
  acceleration} of the wind material.  This effect was discussed for the first
time by \citet{sch1:95} and \citet{sch1:97} and is also detected in our models
\citep{gra2:03}. The clumping factor therefore plays a key role for the dynamics
of WR\,winds.

For the modeling of OB\,star winds, two main approaches are presently followed.
{ In the first one, {\em all} atomic populations are calculated in
  non-LTE (i.e., the statistical equilibrium equations are exactly solved) and
  the radiation field is treated in Sobolev approximation
  \citep{pau1:01,krt1:04}. In the second, a Monte-Carlo technique is employed
  for the radiation transfer which particularly accounts for multiple scattering
  effects due to line-overlaps. The atomic populations of the iron-group
  elements, however, are treated in a modified nebular approximation
  \citep{vin1:99}.}  As mentioned above, both methods are probably not suitable
for the modeling of WR\,winds, because the spectrum formation in WR\,stars is
dominated by non-LTE effects {\em and} line-overlaps.  Despite of the
approximations concerning the atomic populations, all spectral lines are treated
as pure scattering opacities in the Monte-Carlo approach, i.e., they do not
allow for photon leakage from one part of the spectrum to another
\citep[see][]{owo2:99,sim1:04}. This situation may be improved by the new
technique of \citet{luc1:02,luc1:03}. The Sobolev approach, on the other hand,
introduces an artificial distinction between `line radiation' and `background
continuum radiation' that causes problems in case of line overlaps.  A
correction of the continuum radiation field is therefore necessary
\citep[see][]{pau1:94}.  In addition, the {\em first order} Sobolev
approximation breaks down in the acceleration region near the transsonic wind
base -- even for thin stellar winds \citep{sel1:93,owo1:99}.

The hydrodynamic structure of O\,star winds was obtained for the first time by
\citet[][{}\,later on CAK]{cas1:75}, who employed a parametrization of the
radiative acceleration by means of the so-called force multiplier function.
This method has been significantly improved by \citet{pau1:86} and describes
successfully the observed properties of OB\,star winds { \citep[except for dense
  winds where multiple scattering effects become important, see]
  []{lam1:93,pul1:96,vin1:00}. However, due to the parametrization by only two
  parameters, the line force is subject to an averaging process, i.e., the
  hydrodynamic equations are not strictly solved.}  Again, for the highly
structured winds of WR\,stars, this approach is probably not sufficient.  { The
  Monte-Carlo models by \citet{vin1:99} are based on an even less accurate
  method where a consistent {\em global} energy budget is provided for a {\em
    prescribed} velocity field.}  \citet{krt1:04} have recently improved this
situation by a full local solution of the equation of motion {\em and} the
energy equation, i.e., their models provide a consistent velocity and
temperature structure. {Nevertheless, due to the neglect of multiple
  scatterings, these models are limited to weaker winds.}

In the present work we overcome the problems explained above by combining full
non-LTE atmosphere models with the equations of hydrodynamics.  By iteration
between the non-LTE radiation transport and the hydrodynamics we derive
$\dot{M}$, $\velo(r)$, $T(r)$, {\em and} the non-LTE population numbers
consistently with the radiation field in the co-moving frame (CMF), i.e., we
solve the radiation transport, the energy equation, and the equation of motion
simultaneously.  Moreover, clumping is accounted for. The resultant models
describe the conditions in WR\,atmospheres correctly, and provide synthetic
spectra, i.e., they allow for a direct comparison with observational material.
Nevertheless, simplifying assumptions are necessary also in our models.  These
concern especially the assumption of a constant Doppler broadening velocity
throughout the atmosphere and the omission of opacities, partly due to the
neglect of trace elements like Ne, Ar, S, or P and partly due to incompleteness
of the available data { (for ions above ionization stage {\sc ix}, no
  fine-structure data, and no atomic data for elements with $Z > 26$ are
  available)}.

As already mentioned above, the critical-point analysis by \citet{nug1:02}
revealed that the mass-loss of early-type WR\,stars is probably driven by the
`Iron Opacity Bump' which is mainly due to line opacities of the Fe M-shell ions
({\FeIX}--{\sc xvi}). By inclusion of these ions in our models we are able to
check whether the assumptions of \citeauthor{nug1:02} (critical point at high
optical depth and high temperature) are correct, and most importantly, if the
high wind acceleration can be maintained in the outer part of the wind.

{ In the Sect.\,\ref{sec:atmos} of the present work we review our
  standard atmosphere models, followed by a series of test calculations to
  clarify the role of the `Iron Bump' opacities and the clumping factor
  (Sect.\,\ref{sec:testcal}).  Then we describe the combination of our
  atmosphere models with the hydrodynamic equations (Sect.\,\ref{sec:hydromod}),
  and present a self-consistent hydrodynamic atmosphere model that reproduces
  all observed characteristics of an early-type WC\,star.  Finally, we summarize
  our conclusions in Sect.\,\ref{sec:conclusions}.}


\section{Model atmospheres}
\label{sec:atmos}

In the present section we give a short overview of our standard atmosphere
models with {\em prescribed} wind structure, as they are routinely used for the
spectral analysis of WR\,stars
\citep{koe1:92,ham1:92,leu1:94,koe1:95,leu1:96,ham1:98,koe1:02,
  gra1:02,ham1:03}.  In Sect.\,\ref{sec:stdmod} we describe the applied
numerical methods, in Sect.\,\ref{sec:mpar} we review the relevant model
parameters, and in Sect.\,\ref{sec:atoms} we summarize the atomic data utilized
for the calculations.

\subsection{Standard models}
\label{sec:stdmod} 

For our standard models we assume a spherically symmetric, stationary outflow
with a prescribed density and velocity structure.  The model code computes the
radiation field, the atomic populations, and the temperature structure in the
expanding atmosphere. The complete solution comprises three parts, which are
iterated until consistency is obtained: The radiation transport, the
{ statistical equilibrium} equations, and the energy equation.

The radiation transport for the spherically expanding atmosphere is formulated
in the co-moving frame of reference (CMF), neglecting aberration and advection
terms \citep[see][]{mih1:76}.  For a fast solution and a consistent treatment of
the scattering terms we employ the method of variable Eddington factors
\citep{aue1:70}. This means that the moment equations are solved to obtain the
angle-averaged radiation field, and the numerically expensive ray-by-ray
transfer is only calculated from time to time \citep{koe1:02,gra1:02}.  The fast
numerical solution allows a detailed treatment of millions of spectral lines on
a fine frequency grid.

The atomic populations and the electron density are determined from the
equations of statistical equilibrium.  This system of equations is solved in
line with the radiation transport by application of the ALI formalism
\citep[accelerated lambda iteration, see][]{ham2:85,ham1:86}.  Complex model
atoms of He, C, O, Si, and the iron-group are accounted for.  For the inclusion
of millions of iron-group transitions we take advantage of the concept of
super-levels \citep[see][]{gra1:02}.  Furthermore, density inhomogeneities are
accounted for by the assumption of small-scale clumps with a clumping factor $D$
\citep[defined as the inverse of the volume filling factor $f_V$,
see][]{ham1:98}.

The temperature structure is obtained from the assumption of radiative
equilibrium.  In the present work this { constraint equation} is
decoupled from { the equations of statistical equilibrium. It is solved
  by a temperature correction procedure which is based on the Uns\"old-Lucy
  method \citep{uns1:55,luc1:64}, and has been generalized for the application
  in non-LTE models with spherical expansion \citep{ham1:03}.}

\subsection{Model parameters} 
\label{sec:mpar} 
 
The model atmospheres are specified by the luminosity and radius of the stellar
core, the chemical composition of the envelope, and its density and velocity
structure. The basic parameters are: The stellar core radius $R_\star$ at
Rosseland optical depth $\tau_\mathrm{R} = 20$, the stellar temperature
$T_\star$ (related to the luminosity $L_\star$ and the core radius $R_\star$ via
the Stefan Boltzmann law), the chemical composition, the clumping factor $D = 1
/ f_V$, and, if the atmosphere structure is {\em not} calculated
self-consistently, the mass-loss rate $\dot{M}$ and the terminal velocity \vinf.
In the latter case, a $\beta$-type velocity-law of the form
$\velo(r)=\vinf\left(1- R_0/r\right)^\beta$ is assumed, where $R_0$ denotes the
transition point to the hydrostatic domain and $\beta$ is set to one.

At the start of the model calculation, when the density and velocity structure
of the atmosphere are fixed, only the continuum opacities for LTE populations
with a grey temperature structure are calculated.  The value of
$\tau_\mathrm{R}=20$, which specifies the inner boundary of our models,
therefore denotes a `LTE-grey-continuum Rosseland optical depth' of 20. The
`true' Rosseland optical depth, $\tau_{\rm Ross}$, of a converged non-LTE model
with all line opacities included is usually much larger. In our present models
(WC\,stars with solar Fe-abundance) values of $\tau_{\rm Ross} = 50$ are
typically reached at the inner boundary.

Together with the density and velocity structure also the clumping factor $D$ is
prescribed at the start of the model calculation.  For realistic wind models it
is however necessary to allow for a radial variation of $D$.  Especially for our
intended hydrodynamic treatment a smooth wind-structure is favorable at large
optical depth around and below the sonic point. The radial dependence is
therefore specified in the following form.  In the deep layers, where
$\tau_\mathrm{R}$ is greater than a specified value $\tau_1$, a smooth wind with
$D=1$ is assumed. Then the clumping factor is continuously increased until a
fixed maximum value is reached at a specified optical depth $\tau_2$. For
$\tau_\mathrm{R} < \tau_2$, $D$ is kept constant. In the present work, the given
values for $D$ refer to its maximum value with $\tau_1$ set to $0.7$ and
$\tau_2$ to $0.35$.

The parameters $R_\star$, $\dot{M}$, \vinf, and $D$ are connected by the
transformed radius $R_\mathrm{t} \propto R_\star ( \vinf / \sqrt{D}\dot{M}
)^{2/3}$.  Models with the same $R_\mathrm{t}$ show almost identical line
equivalent widths \citep{sch1:89}. This invariance holds because the line
emission in WR\,stars is dominated by recombination processes
\citep[see][]{ham1:98}.

\subsection{Atomic data}
\label{sec:atoms} 

\begin{table}[t]
\begin{tabular}{llllll}
\hline \hline 
\rule{0cm}{2.2ex} Ion & Levels && Ion & Super-levels & Sub-levels\\
\hline
\rule{0cm}{2.2ex}\HeI& 17            &&\FeII&   1  & 1\\
\HeII& 16                            &&\FeIII&  8  & 14188\\
\HeIII& 1                            &&\FeIV&   18 & 30122\\
\CI&2                                &&\FeV&    19 & 19804\\
\CII& 32                             &&\FeVI&   18 & 15155\\
\CIII& 40                            &&\FeVII&  16 & 11867\\
\CIV& 54                             &&\FeVIII& 10 & 8669\\
\CV& 1                               &&\FeIX&   11 & 12366\\
\OII& 3                              &&\FeX&    9  & 761\\      
\OIII& 33                            &&\FeXI&   9  & 1143\\
\OIV& 25                             &&\FeXII&  9  & 894\\
\OV& 36                              &&\FeXIII& 9  & 797\\
\OVI& 15                             &&\FeXIV&  9  & 589\\
\OVII& 1                             &&\FeXV&   7  & 253\\
\SiIII& 10                           &&\FeXVI&  6  & 31\\
\SiIV& 7                             &&\FeXVII& 1  & 1\\
\SiV& 1 &&&\\
\hline
\end{tabular} 
\caption{Summary of the extended model atom. The iron-group ions (Fe)
are described by a relatively small number of super-levels, each 
representing a large number of true atomic energy levels (sub-levels). 
In the present work we also apply a smaller model atom, where
the iron-group ions are restricted to {\FeII}--{\FeX} with {\FeX}
represented by one auxiliary level.}
\label{tab:modelatom}
\end{table}

The chemical composition of the model atmospheres is given by mass fractions
$X_{\rm He}$, $X_{\rm C}$, $X_{\rm O}$, $X_{\rm Si}$ and $X_{\rm Fe}$ of helium,
carbon, oxygen, silicon and iron-group elements. Compared to \citet{gra1:02} the
atomic data are extended by the ions {\FeX}--{\sc xvii} and by detailed
photoionization cross sections for C and O that include dielectronic resonances.
Two types of model atoms, with and without the extended iron model, are applied
in the present work. The model atoms are summarized in
Table\,\ref{tab:modelatom}.  They contain ionization stages {\HeI}--{\HeIII},
{\CI}--{\CV}, {\OII}--{\OVII}, {\SiIII}--{\SiV}, and {\FeII}--{\FeX} or
{\FeII}--{\FeXVII}.  Except for {\HeI} and {\SiIII}, the lowest and highest
ionization stages are restricted to few auxiliary levels.

The atomic level \& line data are compiled from the following sources.
Oscillator strengths are taken from the Opacity Project
\citep{hum1:88,sea1:87,sea1:92,cun1:92,sea1:94,the1:95}, level energies from
Kurucz's CD-ROM No.\,23 \citep{kur1:95}, iron group data (ionization stages
{II}--{IX}) from Kurucz CD-ROM No.\,20--22 \citep[see][]{kur1:91,kur1:02}, and
the iron M-shell data ({\FeX}--{\FeXVII}) from the Opacity Project.  The
iron-group elements from Sc to Ni are comprised into one generic atom.  This
model atom is of course dominated by Fe due its high relative abundance.  The
extension of the generic ion from the stages {\FeX} to {\FeXVII} is restricted
to pure Fe because the Opacity Project does not provide data for elements with
$Z>26$.

The emission line spectra of WR\,stars are chiefly formed in line emission
cascades, which follow recombination to excited energy levels. A realistic
treatment of photoionization and recombination processes is therefore mandatory.
Especially for the ionization of carbon and oxygen ions {\CI}--{\CIII} and
{\OI}--{\OV}, the excitation of auto-ionizing, dielectronic resonances close to
the ionization edge provides an important reaction cannel.  In our previous
work, photoionization cross sections were obtained by low-order fits to Opacity
Project data, and the dielectronic transitions were treated separately by the
approach of \citet{mih1:73}.  This distinction between dielectronic and `pure'
photoionization processes is, however, highly questionable.  The resonances
actually form complex features on the total cross section, which cannot be
eliminated by a fitting process. In the present work we avoid these problems by
taking the ionization cross sections for C and O by \citet{nah1:97} and
\citet{nah1:99} as they are, i.e., they are only smoothed with a Doppler
profile.  Effects like the successive saturation of the dielectronic features
when a continuum becomes optically thick, are therefore modeled in a realistic
manner.  The numerical cost of this treatment is, however, rather high because
the continuum opacities must be provided on a fine frequency grid to resolve the
dielectronic features.  Further disadvantages may be due to the slightly
incorrect wavelengths in the data source. Moreover, the cross sections are only
available in a form where the transitions from one lower level to several states
of the upper ion are combined. Photoionizations therefore always lead into the
ground state of the upper ion.

\section{Test calculations}
\label{sec:testcal}
 
 \begin{table}[] 
  \begin{tabular}{llllll} 
    \hline \hline 
    \rule{0cm}{2.2ex}Model & GKH02 & A,B & C & D & Hydro \\
    \hline
    \rule{0cm}{2.2ex}$L_\star /  L_\odot $ & $10^{5.45}$ & $10^{5.45}$ & $10^{5.45}$ & $10^{5.45}$ & $10^{5.45}$ \\
    $R_\star / R_\odot$ & 2.455 & 2.455 & 0.905 & 0.905 & 0.905 \\ 
    $T_\star / \mathrm{kK}$ & 85 & 85 & 140 & 140 & 140 \\ 
    $M_\star / M_\odot$ & 13.63 & 13.63 & 13.63 & 13.63 & 13.63 \\
    \hline 
    \rule{0cm}{2.2ex}$\dot{M} / (\msunpyr)$ & $10^{-4.90}$ & $10^{-4.86}$ & $10^{-4.86}$ & $10^{-5.21}$ & $10^{-5.14}$ \\
    $R_\mathrm{t} / R_\odot$ & 4.132 & 3.683 & 1.358 & 1.358 & 1.210 \\ 
    $D$ & 10 & 10 & 10 & 50 & 50 \\
    $\vinf / (\kms)$ & 2200 & 2020 & 2020 & 2020 & 2010 \\ 
    \hline 
    \rule{0cm}{2.2ex}$X_\mathrm{He}$ & 0.51 & 0.33 & 0.33 & 0.33 & 0.33 \\
    $X_\mathrm{C}$  & 0.45 & 0.60 & 0.60 & 0.60 & 0.60 \\
    $X_\mathrm{O}$  & 0.04 & 0.06 & 0.06 & 0.06 & 0.06 \\
    $X_\mathrm{Si}/10^{-3}$ & $0.8$ & $0.8$ & $0.8$ & $0.8$ & $0.8$ \\
    $X_\mathrm{Fe}/10^{-3}$ & $1.6$ & $1.6$ & $1.6$ & $1.6$ & $1.6$ \\
    \hline
    \rule{0cm}{2.2ex}$Q$ & 0.42 & 0.47 & 0.71 & 1.05 & 1.00 \\
                  $\eta$ & 4.83 & 4.87 & 4.87 & 2.17 & 2.54 \\
    \hline
  \end{tabular} 
  \caption{Model parameters for our different test models (A,B,C, and D)
  and the hydrodynamic model (Hydro) compared to the values derived for
  WR\,111 by \citep{gra1:02}.
 {\em Stellar core parameters}: 
 Stellar luminosity $L_\star$, radius $R_\star$,
 temperature $T_\star$, and mass $M_\star$
 (relevant only for the hydrodynamic model). 
 {\em Wind parameters}: 
 Mass-loss rate $\dot{M}$ with the corresponding transformed radius $R_{\rm t}$,
 clumping factor $D$, and terminal wind velocity \vinf. 
 {\em Atmospheric abundances}: 
 Mass fractions of helium, carbon, oxygen, silicon and iron
 $X_{\rm He}$, $X_{\rm C}$, $X_{\rm O}$, $X_{\rm Si}$, and $X_{\rm Fe}$.
 {\em Wind efficiency}: 
 Resultant work-ratio $Q=W_\mathrm{wind}/L_\mathrm{wind}$
 according to Eqs.\,(\ref{eq:wwind}) and (\ref{eq:lwind}),
 and wind momentum efficiency
 $\eta = L_\star / c\dot{M} \vinf$.}
  \label{tab:parameters} 
\end{table}

{ As a preparation for our intended hydrodynamic modeling, we
  investigate in the present section the effect of the additional Fe opacities
  on the radiative acceleration, and determine the range of stellar parameters
  that actually allows to maintain a WR-type wind.} For this purpose we employ
models with a prescribed $\beta$-type velocity law and similar parameters as the
WC\,5 prototype WR\,111, which has already been analyzed in detail by
\citet{hil1:99} and \citet{gra1:02}. Although the modeling of WC\,stars is a
rather complex task, these objects offer a good test case for our hydrodynamic
models.  Due to their almost homogeneous chemical composition their mass can be
reliably derived from the theoretical mass-luminosity relation e.g.\ by
\citet{lan1:89}, i.e., the number of free parameters is reduced by one.

In Sect.\,\ref{sec:fetest} we identify the line opacities of the Fe M-shell ions
as the main contributors to the wind acceleration in deep atmospheric layers,
and show that the WR-type mass-loss can only be maintained for rather high
stellar core temperatures (around $T_\star=140\,\mathrm{kK}$). Furthermore, we
point out the influence of the clumping factor on the radiative acceleration.
In Sect.\,\ref{sec:clumptest} we demonstrate the reaction of the optical
\OVI-features on the radial dependence of the clumping factor, and show that
they are a possible diagnostic tool for the location where the wind clumping
sets in.

\subsection{The effect of the Fe M-shell opacities}
\label{sec:fetest}

To investigate the effect of the additional Fe opacities we start with a similar
model (i.e.\ with similar parameters and atomic data) as in our previous
analysis of WR\,111 \citep{gra1:02}.  The model parameters are adapted in a way
to facilitate the intended hydrodynamic treatment (slightly higher C and O
abundances) and to allow for an easy comparison to the final model in
Sect.\,\ref{sec:hydrocal} ($\vinf \rightarrow 2020\,\kms$). Moreover, the
mass-loss rate is adjusted to reproduce the observed \CIII/\CIV\ line strengths.
Note that the effective wind density is enhanced in our present models due to
the improved treatment of dielectronic recombination (see
Sect.\,\ref{sec:atoms}). In Table\,\ref{tab:parameters}, the parameters of this
first test model (Model~A) are listed together with the values previously
derived for WR\,111 (GKH02).

\begin{figure}[]
  \parbox[b]{0.49\textwidth}{\includegraphics[height=0.34\textwidth]{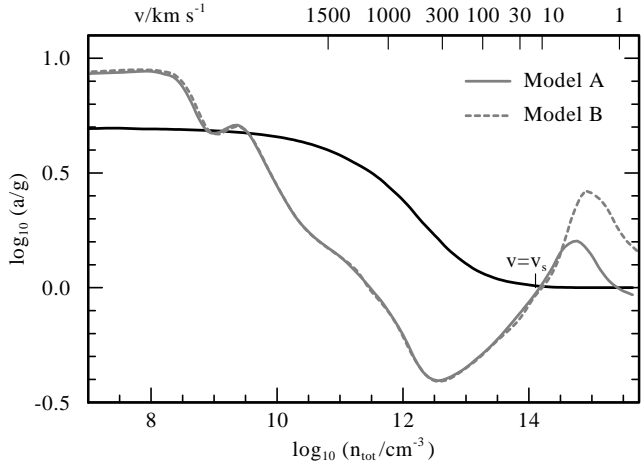}}
\caption{
  Test models A and B ($T_\star = 85\,\mathrm{kK}$) with {\em prescribed}
  velocity field: Wind acceleration in units of the local gravity $g(r)$ {\em
    vs}.\ atomic density as depth index.  The mechanical + gravitational
  acceleration as implied by the adopted $\beta$-law (black) is compared to the
  wind acceleration due to radiative + gas pressure.  Whereas Model~A (grey)
  comprises the iron group ions \FeII--\FeX, Model~B (grey dashed) is extended
  by the full iron M-shell up to \FeXVII.}
\label{fig:acc1}
\end{figure}

\begin{figure}[]
  \parbox[b]{0.49\textwidth}{\includegraphics[height=0.34\textwidth]{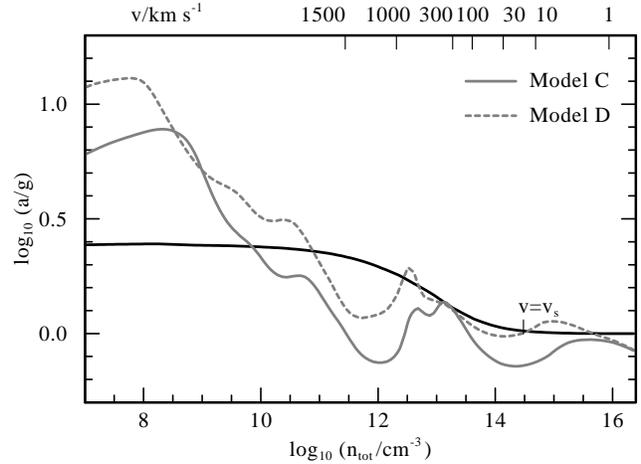}}
\caption{
  `Hot' test models C and D ($T_\star = 140\,\mathrm{kK}$) with {\em prescribed}
  velocity field: Wind acceleration as in Fig.\,\ref{fig:acc1} but for models
  with $T_\star = 140\,\mathrm{kK}$. For Model~D (grey, dashed) the clumping
  factor has been increased from $D=10 \rightarrow 50$ and the mass-loss rate
  has been correspondingly reduced by $\sqrt{50/10}$.  For this model a globally
  consistent energy budget with $Q = 1.05$ is obtained whereas Model~C only
  provides a fraction of $Q=0.71$ of the wind energy.}
\label{fig:acc2}
\end{figure}

The radiative acceleration as obtained from Model~A is shown in
Fig.\,\ref{fig:acc1}. It is compared to the wind acceleration according to the
prescribed $\beta$-type velocity law. In analogy to previous works
\citep{gra1:00,her1:01,gra1:02,hil1:03}, the obtained radiative force exceeds
the force needed to accelerate the wind in the outer layers, but fails at large
optical depth where the mass-loss is initiated.  For the work ratio $Q$ between
the work performed by the radiation field and the gas expansion on the wind
material
\begin{equation}
\label{eq:wwind}
W_\mathrm{wind} = \dot{M}\int_{R_\star}^\infty \left( a_{\rm rad}
 - \frac{1}{\rho}\frac{\ddif p}{\ddif r}\right) \dif r
\end{equation}
and the prescribed mechanical wind luminosity
\begin{equation}
\label{eq:lwind}
L_{\rm wind} = \dot{M} \left( \frac{1}{2}\,\vinf^2 +  \frac{M_\star G}{R_\star} 
\right),
\end{equation}
we obtain a value of $Q = W_{\rm wind}/L_{\rm wind} = 0.47$, i.e.\ the model
provides about half of the global energy budget of the wind.

\begin{figure*}[t]
  \parbox[b]{0.66\textwidth}{\includegraphics[width=0.63\textwidth]{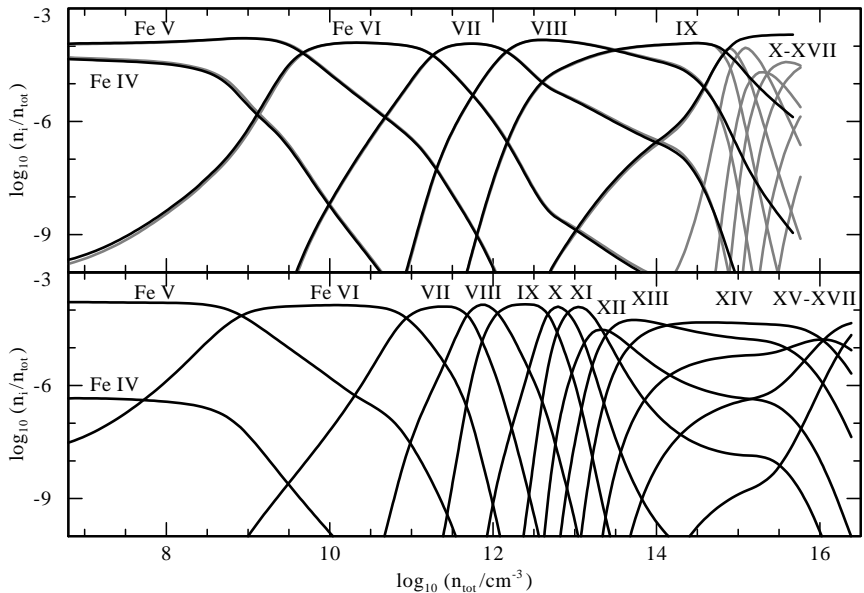}}
  \parbox[b]{0.33\textwidth}{
\caption []{Test calculations:
  Ground-state populations of the included iron-group ions {\em vs}.  atomic
  density as depth index for models A, B, and C.  {\em Top panel}: The `cool'
  models A (black) and B (grey) show ionization stages up to \FeXIV. The model
  atom for Model~A which only includes ions up to \FeX\ is clearly insufficient.
  For both models, the M-shell ions (\FeIX--\FeXVII) are strongly concentrated
  towards the inner boundary. {\em Bottom panel}: The `hot' Model~C shows
  M-shell ions already at lower densities.  Ionization stages up to \FeXVI\ are
  excited.}
  \label{fig:ironpop}
}
\end{figure*}

The reason for the lack of force in the acceleration region becomes obvious when
the atomic populations of Model~A are inspected.  The top panel of
Fig.\,\ref{fig:ironpop} shows the Fe-ionization structure of Model~A (black).
In the inner part of the atmosphere, Fe is ionized up to the highest ionization
stage considered in the model atom, i.e., in this region only opacities from
subordinate levels contribute to the radiative force. The lack of force may
therefore be attributed to a lack of sufficiently high ionization stages
considered in the model atom. For C and O the situation is similar, however,
these elements are complete up to their He-like ions. Fe is only considered up
to the second M-shell ion, \FeX.  Because the ionization potentials of the
M-shell ions are increasing only in small steps from 236\,eV for \FeIX\ to
489\,eV for \FeXVI\ before a big step of 1266\,eV follows for \FeXVII, one may
expect that at least a few more M-shell ions could be excited if they were
present in the model atom.

That this is indeed the case is indicated by the grey curves in
Fig.\,\ref{fig:ironpop}. They show the populations of Model~B which has the same
parameters as Model~A but an extended model atom, including all M-shell ions up
to \FeXVII\ (see Table\,\ref{tab:modelatom}). In this model, ions up to \FeXIV\ 
are excited.  However, an inspection of Fig.\,\ref{fig:acc1} shows that the
radiative force due to the additional ions does not solve our problem with the
wind acceleration. $a_\mathrm{rad}$ is only increased in the hydrostatic layers,
where it now considerably exceeds the gravitational attraction, i.e., no
hydrostatic equilibrium is possible anymore.  The deficit in the acceleration
region remains.

The apparent solution of this problem is an increase of $T_\star$, to push the
temperature regime where the M-shell ions are excited farther out into the
acceleration region of the wind. For a typical WC\,model, the emergent spectrum
is only marginally changed by this operation because the continuum becomes
optically thick already at high wind velocities, i.e., the location of the
photosphere is determined by the velocity structure of the wind rather than the
size of the stellar core \citep[see][]{ham2:03}. In the deeper layers however,
where our problem with the radiative acceleration occurs, the temperature scales
with $T_\star$ and the ionization structure may be changed considerably.

In the lower panel of Fig.\,\ref{fig:ironpop} we show the Fe-populations of
Model~C where the stellar radius $R_\star$ has been correspondingly decreased so
that $T_\star=140\,\mbox{kK}$.  It is clearly visible that the bulk of the
M-shell ions is now shifted towards much lower densities, and ions up to \FeXVI\ 
are exited.  As demonstrated in Fig.\,\ref{fig:acc2}, the radiative acceleration
on these ions works exactly in the intended way, and fills the previous gap in
the acceleration region.  Moreover, the work ratio $Q$ of this model has
improved to a value of 0.71.  Nevertheless, below and directly above the
apparent opacity bump, the radiative acceleration still lies slightly below the
gravitational attraction. Whereas the deficit in the deeper layers, below
$\log(n_\mathrm{tot}/\mathrm{cm}^{-1}) = 13.5$, would be balanced by the gas
pressure in a hydrodynamic model, the lack of force around
$\log(n_\mathrm{tot}/\mathrm{cm}^{-1}) = 12$, would presumably remain.  The
latter, however, is located so far outside in the wind that changes in the
recombination rates due to clumping can considerably affect the wind dynamics.

This is demonstrated by Model~D in Fig.\,\ref{fig:acc2}, where the clumping
factor has been increased from $D=10$ to a value of $50$. The mass-loss rate has
been correspondingly reduced by a factor of $\sqrt{50/10}$ in order to maintain
the transformed radius $R_{\rm t}$ and thereby the strength of the emergent
emission line radiation (see Sect.\,\ref{sec:mpar}).  For this model a work
ratio of 1.05 is obtained, i.e., it has a nearly consistent global energy
budget. With respect to Model~C, the radiative acceleration is partially
enhanced by up to 0.3\,dex.  The influence of the clumping factor and its exact
dependence on the radius is therefore crucial for the dynamics of WR\,winds.

{ The lack of force in Model~C is presumably due to the omission of
  trace elements like Si, P, S, Ne, and especially Ar in our calculations
  \citep[see][]{hil1:03}, i.e., the high clumping factors in Model~D are
  possibly only needed as a compensation for this deficit.}  Nevertheless, we
use Model~D as the start model for our hydrodynamic modeling in
Sect.\,\ref{sec:hydrocal}, and admit that we might overestimate the clumping
factor in order to achieve a consistent solution.

Interestingly, with an electron temperature of $140$--$200\,\mbox{kK}$ in the
acceleration region, our Model~D basically matches the conditions claimed by
\citet{nug1:02} for the initiation of an optically thick wind.  However, our
models indicate that the acceleration of the wind above the sonic point is only
possible for very high core temperatures of $T_\star \approx 140\,\mathrm{kK}$.
This is in contradiction to \citeauthor{nug1:02} who assumed much larger radii,
in accordance with current spectral analyses { \citep[see
  also][]{lam1:02}}.  These typically provide values of $T_\star <
90\,\mathrm{kK}$ \citep{koe1:95,gra1:98,gra1:02,cro1:02}.  Our present value, on
the other hand, is in accordance with stellar structure calculations which also
show high core temperatures \citep[120--130\,kK, see][]{lan1:89,sch1:96}.  This
may imply that, due to the obscuration of the stellar core by the wind opacity,
spectral analyses tend to overestimate the stellar radius, i.e., the large
derived radii do actually reflect a very extended wind-structure rather than a
large core size.

\subsection{The radial dependence of the clumping factor}
\label{sec:clumptest}

As explained in Sect.\,\ref{sec:mpar}, the radial dependence of the clumping
factor is fixed at the start of our model calculations.  For the models in
Sects.\,\ref{sec:fetest} and \ref{sec:hydrocal}, the clumping factor is assumed
to increase from a homogeneous wind ($D=1$) for optical depths larger than
$\tau_1=0.7$ to a specified value ($D=10$ or 50) for optical depths smaller than
$\tau_2=0.35$.  As demonstrated in the previous section, clumping has a strong
impact on the radiative acceleration, i.e., the results of our hydrodynamic
modeling will depend on the detailed choice of $\tau_1$ and $\tau_2$.

For a justification of our present choice we have calculated test models similar
to Model~D but with different values of $\tau_1$ and $\tau_2$.  For Model~D1 the
clumping factor is assumed to be constant throughout the atmosphere, whereas for
Model~D2, $\tau_1$ and $\tau_2$ have been decreased ($\tau_1=0.3$, $\tau_2=0.1$)
so that the onset of clumping is pushed farther out in the wind.

In the emergent spectrum these changes especially show up in the narrow
\OVI-doublet at 3811/34\,\AA. In previous works
\citep[e.g.][]{hil1:99,gra1:02,cro1:02} these lines could not be reproduced,
which is remarkable because of the relatively simple structure of the {\OVI}
ion.  That we are now able to reproduce {\OVI} features of the observed strength
is due to the combination of the high stellar temperature and the additional
blanketing by the Fe M-shell opacities. \OVI\ consequently helps to trace
regions relatively deep inside the envelope at temperatures around 100\,kK.

\begin{figure}[tp!]
  \parbox[]{0.49\textwidth}{\includegraphics[width=0.45\textwidth]{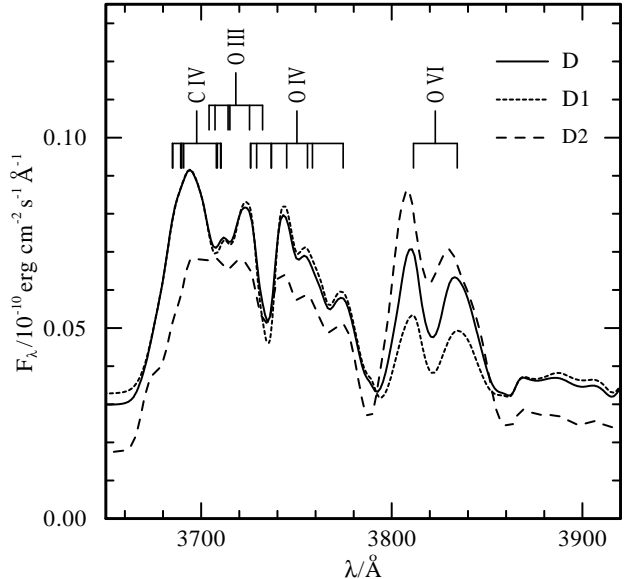}}
\caption []{
  Model spectra around \OVI\,3811/34 for different clumping structures. For
  Model~D our standard approach is used (see text). For Model~D2 clumping is
  initiated farther out in the wind whereas for Model~D1 a constant clumping
  factor is assumed.  The \OVI\ emissions react strongly on the clumping
  structure.  The farther out clumping is initiated the stronger becomes \OVI.
  For Model~D2 also the continuum-level is affected.}
\label{fig:ovi}
\end{figure}

As demonstrated in Fig.\,\ref{fig:ovi}, \OVI\,3811 reacts strongly on the
performed changes of the clumping factor. In our test, Model~D shows slightly
stronger emissions than observed. Model~D1, which is fully clumped, provides
much weaker lines, whereas for Model~D2, where the clumping sets in farther out,
not only \OVI\ is enhanced, but also the flux level of the continuum is
affected.  Although the strength of the {\OVI} lines will depend on the final
model parameters (see next section), the present test shows that \OVI\ is a good
diagnostic tool for the clumping structure in deep atmospheric layers.

\section{The wind hydrodynamics}
\label{sec:hydromod}

In this section we describe the combination of our model atmospheres with the
equations of hydrodynamics. The resultant atmosphere models provide a
self-consistent wind structure, i.e., for given stellar core parameters the
non-LTE population numbers {\em and} the atmosphere structure are derived
consistently with the radiation field in the co-moving frame of reference (CMF).
The solution is obtained by iteration between the non-LTE radiation transport
and the hydrodynamics. Thereby, the numerical effort is considerably enhanced
with respect to the standard models.  In Sect.\,\ref{sec:arad} we discuss the
parametrization of the radiative acceleration, which forms the link between the
radiation transport and the hydrodynamics. The solution of the hydrodynamic
equations is explained in Sect.\,\ref{sec:hydro}.

\subsection{The radiative acceleration}
\label{sec:arad}

From the solution of the moment equations in the radiation transport, we obtain
the Eddington flux $H_\nu$ on a fine frequency grid in the CMF.  In combination
with the opacity $\kappa_\nu$, which is given on the same grid, it is thus
possible to derive the radiative acceleration on the wind material by direct
integration
\begin{equation}
\label{eq:arad}
a_{\rm rad} = \frac{1}{\rho}\,\frac{4\pi}{c} \int_0^{\infty} \kappa_\nu
H_\nu\,\dint \nu.
\end{equation}

Due to Doppler shifts, however, the radiative acceleration on {\em spectral
  lines} depends strongly on the velocity gradient $\velo'$. The
radiative-acoustic wave modes arising from this dependence are changing the
characteristic properties of the gas flow \citep{abb1:80}.  For the application
in our hydrodynamics it is therefore necessary to find a suitable
parametrization of $a_{\rm rad}$ which accounts for this effect.

In the CAK approach \citep{cas1:75} the acceleration on spectral lines is
written in the form
\begin{equation}
  \label{eq:acak}
  a_{\rm lines}
  = \frac{C}{r^2} \left( \frac{1}{\rho} \frac{\dif \velo}{\dif r} \right)^\alpha
  \equiv \frac{C}{r^2} \left( \frac{\dif \velo}{\dif \tau} \right)^\alpha,
\end{equation}
where we have used our own definition $\dif \tau \equiv \rho\,\dif r$. In this
formulation the parameter $\alpha$ reflects the distribution of line strengths
in the relevant wavelength range, and is usually found to be { in the
  range 0.6--0.7}.  In contrast to this, the radiative acceleration in our
models, as calculated from Eq.\,(\ref{eq:arad}), cannot be described by two
parameters. To obtain consistent wind models we have to allow for {\em depth
  dependent} parameters $C$ and $\alpha$. By this, however, the solution
topology due to the non-linearity in $\velo'$ becomes much more complicated. To
avoid numerical problems we therefore utilize a linearized form of
Eq.\,(\ref{eq:acak}) in our calculations.

Let us first assume that $a_{\rm lines}$ is given for a {\em fixed} velocity
field $\velo_0(r)$. Then a first order expansion of Eq.\,(\ref{eq:acak}) with
respect to $\velo'$ gives
\begin{equation}
  \label{eq:aapp}
  a_{\rm lines} \approx
  (1-\alpha) \frac{C}{r^2} \left( \frac{\dif \velo_0}{\dif \tau} \right)^\alpha
  + \alpha  \frac{C}{r^2} \left( \frac{\dif \velo_0}{\dif \tau} \right)^{\alpha - 1}
        \left( \frac{\dif \velo}{\dif \tau} \right).
\end{equation}
Here the first term $\propto 1/r^2$ behaves as expected for optically thin lines
(note that $\velo_0$ is fixed), and the second term $\propto \frac{{\rm d}\velo}
{{\rm d}r}/\rho r^2$ reflects the behavior of optically thick lines due to
Doppler shifts, i.e., the parameter $\alpha$ indicates the fraction of the line
force which is due to optically thick lines
{\citep[see][]{gay1:95,pul1:00}}.

For a realistic representation in the hydrodynamics, $a_{\rm rad}$ from
Eq.\,(\ref{eq:arad}) must be disentangled into its different contributions.  As
a first step, the line acceleration $a_{\rm lines}$ must be separated from the
acceleration on continua $a_{\rm cont}$. Then the ratio between optically thin
and optically thick lines must be determined, so that $a_{\rm rad}$ can be
written as $a_{\rm cont} + a_{\rm thin} + a_{\rm thick}$.  $a_{\rm thin}$ and
$a_{\rm thick}$ correspond to the first and second term in Eq.\,(\ref{eq:aapp}),
i.e., they must be parametrized in a form where $a_{\rm thin} \propto 1/r^2$ and
$a_{\rm thick} \propto \frac{{\rm d}\velo}{{\rm d}\tau}/r^2$.  Because the
acceleration on continua is also independent from the velocity gradient, $a_{\rm
  cont}$ can be treated analogously to $a_{\rm thin}$.  Finally, $a_{\rm rad}$
can be written in the form
\begin{equation}
 \label{eq:apar}
 a_{\rm rad}  = \left(a_{\rm cont} + a_{\rm thin}\right) + a_{\rm thick} \equiv
 \frac{M G\,\Gamma(\tau)}{r^2}
 + \frac{k(\tau)}{r^2} \left( \frac{\dif \velo}{\dif \tau} \right),
\end{equation}
where $\Gamma(\tau)$ includes the contributions from continua and optically thin
lines, and $k(\tau)$ describes the contribution from optically thick lines.

The first step, the separation of $a_{\rm cont}$ and $a_{\rm lines}$, is easily
done by dividing the total opacity $\kappa_\nu$ under the integral in
Eq.\,(\ref{eq:arad}) into the parts due to continua and lines only.  Then the
dependence of $a_{\rm lines}$ on $\velo'$ is investigated to determine $a_{\rm
  thin}$ and $a_{\rm thick}$.  For this purpose the velocity field $\velo(r)$ in
the radiation transport is varied, i.e., the radiation field is calculated twice
for the same density structure and the same atomic populations but for different
velocity distributions $\velo(r)$. In practice, a variation of 10\% is applied
to $\velo(r)$ and $\velo'(r)$.  The variation of $\velo'$ then causes changes in
the derived value of $a_{\rm lines}$ {\em and} the total radiative flux.  After
normalizing the result with respect to the flux, we obtain a measure for the
reaction of $a_{\rm lines}$ on the variation of $\velo'$ alone.  Now the
fraction of the line force due to optically thick lines (which is equivalent to
the force multiplier parameter $\alpha$) can be calculated directly from
Eq.\,(\ref{eq:aapp}) for each depth index.

The physical meaning of $\alpha$ derived on this way is different from the
standard approach, where $\alpha$ is { interpreted as a characteristic
  material property due to the distribution of line strengths.  The value
  derived here is an {\em effective} value which reflects} the reaction of
$a_{\rm lines}$ on $\velo'$ {\em including} all important radiative effects like
multiple scattering, line overlaps, ionization effects, the exact
angle-dependence of the radiation field, or the shortening of the photon mean
free path at large optical depth, which also have partly been introduced in the
standard CAK approach \citep[see][]{abb1:82,pau1:86,pul1:00}.

The parametrization of $a_{\rm rad}$ in Eq.\,(\ref{eq:apar}) is fundamentally
different from the CAK approach (Eq.\,\ref{eq:acak}) which is non-linear in
$\velo'$. Our linearization helps to avoid numerical problems due to multiple
solutions.  However, in the CAK approach the non-linearity is prerequisite for
the self-consistent calculation of the mass-loss rate.  For our present
calculations, we therefore need an additional constraint.  { Because
  $\Gamma$ and $k$ are continuously updated, the overall nonlinear behavior of
  the line force is nevertheless retained during the iterative solution.}

\subsection{Solution of the hydrodynamic equations}
\label{sec:hydro}

The self-consistent hydrodynamic atmosphere models are obtained by iteration
between the non-LTE radiation transport as described in Sect.\,\ref{sec:stdmod},
and the solution of the hydrodynamic equations.  The radiation transport
provides the radiative acceleration according to Eq.\,(\ref{eq:apar}) and the
temperature structure, namely $\Gamma(\tau)$, $k(\tau)$, and $T(\tau)$.  Based
on this, a hydrodynamic solution for the density and velocity structure,
$\rho(r)$ and $\velo(r)$, is obtained.  $\rho$ and $\velo$ are then fed back
into the radiation transport, and the procedure is repeated until convergence.

As already indicated above, $T(\tau)$ as well as $\Gamma(\tau)$ and $k(\tau)$
are fixed on the $\tau$-grid, where $\tau$ is given by our simple definition
$\ddif\tau = \rho\,\ddif r$ from Eq.\,(\ref{eq:acak}).  By this, the energy
equation is implicitly included in the hydrodynamics (at large optical depth $T$
is coupled to $\tau_\mathrm{Ross}$).  This is particularly important for the
optically thick winds of WR\,stars where a violation of radiative equilibrium
results in extreme deviations from flux constancy, which would spoil the whole
iteration process.

To maintain radiative equilibrium it is further necessary that the {\em total}
optical depth of the atmosphere remains unchanged when a new hydrodynamic
structure is calculated. As explained in the following, this additional
constraint is exactly what we need to close the system of hydrodynamic
equations, i.e., by the inclusion of the energy equation we gain the additional
constraint which is required due to our linearization of $a_\mathrm{rad}$ with
respect to $\velo'$.  We incorporate the conservation of optical depth into the
hydrodynamics by choosing $\tau$ as the integration variable.  Due to our simple
definition of $\tau$, however, we neglect the the actual $\rho$-dependence of
flux-mean opacity.  Nevertheless, this approach turned out to be numerically
stable {\em and} efficient enough for our purposes.

With $a_{\rm rad}$ from Eq.\,(\ref{eq:apar}), the equation of motion for a
stationary, spherically expanding gas flow reads
\begin{equation}
\label{eq:mor}
\velo \frac{\ddif \velo}{\ddif r} = -\frac{M G\,(1- \Gamma)}{r^2}
-\frac{1}{\rho}\frac{\ddif p}{\ddif r} +\frac{k}{r^2} \left(\frac{1}{\rho}\frac{\ddif \velo}{\ddif r}\right),
\end{equation}
where $M$ denotes the stellar mass, $G$ the gravitational constant, $\rho$ the
gas density, and $p$ the gas pressure.  Formulating $p$ in terms of the
isothermal sound speed $a=\sqrt{p/\rho}$, replacing $\ddif r$ by $\ddif \tau /
\rho$, and applying the equation of continuity
\begin{equation}
  \label{eq:cont}
  \dot{M} = 4\pi\,\rho \velo r^2,
\end{equation}
we obtain the equation of motion in the form
\begin{equation}
\label{eq:motion}
  \frac{\ddif \velo}{\ddif \tau}
  \left[\left(1-\frac{{a}^2}{\velo^2}\right)\,\frac{\dot{M}}{4\pi} - k\right] \\
  = - M G\,(1-\Gamma) + 2\,a^2r - \frac{\dot{M}}{4\pi \velo}
  \frac{\dint a^2}{\ddif \tau},
\end{equation}
where $\Gamma(\tau)$, $k(\tau)$, and $a(\tau)=\sqrt{{\cal R}T(\tau)/\mu}$ are
taken from the preceding non-LTE model calculation.  This equation must be
solved in line with $r(\tau)$ which is given by our definition of $\tau$
\begin{equation}
  \label{eq:rad}
  \frac{\ddif r}{\ddif \tau} = \frac{1}{\rho} = \frac{4\pi \velo r^2}{\dot{M}}.
\end{equation}

For given boundary values $\velo_{\rm b}$ and $r_{\rm b}$, and a given mass-loss
rate $\dot{M}$, Eqs.\,(\ref{eq:motion}) and (\ref{eq:rad}) can be integrated
directly over $\tau$. The outer boundary of our model atmosphere is located at a
prescribed radius $r_\mathrm{b}=R_\mathrm{max}$. At this point $\tau$ can be
arbitrarily set to $\tau_{\rm b}=0$. The inner boundary value $\tau_\star$ at
the stellar core radius $R_\star$ is taken from the previous non-LTE solution.
By these two conditions, namely $r(\tau_\star)=R_\star$ and $r(0)=R_{\rm max}$,
we ensure that the optical depth of our atmosphere model is conserved.  A third
condition is obtained at the critical point of the equation of motion, where the
term in square brackets in Eq.\,(\ref{eq:motion}) becomes zero. In analogy to
the CAK approach, this point marks the position where the flow speed reaches the
value of the fast radiative-acoustic wave mode \citep{abb1:80}. At this point
also the right hand side of Eq.\,(\ref{eq:motion}) must be zero for $\ddif \velo
/\ddif \tau$ to be defined.

After all we have three conditions to obtain two boundary values $\velo_{\rm b}$
and $r_{\rm b}$, and the mass-loss rate $\dot{M}$.  For the starting point of
the numerical integration we choose the outer boundary, which is located at
radius $R_{\rm max}$ with $\tau=0$. By this, the first of our three conditions
($r(0)=R_{\rm max}$) is fulfilled, and the values $\velo(0) \equiv \vinf$ and
$\dot{M}$ remain to be determined.  As start values for $\vinf$ and $\dot{M}$ we
take the values from the previous model iteration. Then the numerical
integration of Eqs.\,(\ref{eq:motion}) and (\ref{eq:rad}) is performed, until
the critical point is reached.  If the critical condition is not fulfilled,
$\dot{M}$ is adjusted and the integration is repeated until this is the case.
Now the numerical integration can be continued through the critical point to the
inner boundary at $\tau=\tau_\star$, where our last condition
$r(\tau_\star)=R_\star$ must be fulfilled. This is achieved by an adjustment of
$\vinf$, and repetition of the previous procedure until $r(\tau_\star)=R_\star$.

Instead of using a pure numerical algorithm we employ relatively simple rules
for the adjustment of $\vinf$ and $\dot{M}$.  At the critical point, the new
value for $\dot{M}$ can be read off directly from Eq.\,(\ref{eq:motion}).
$\vinf$, on the other hand, can be adjusted by the well-known result from the
CAK theory that the terminal velocity of a radiatively driven wind scales with
the escape velocity at the stellar surface, i.e., if $r(\tau_\star) \ne R_\star$
we scale the terminal velocity by $\velo_\mathrm{esc}(R_\star) /
\velo_\mathrm{esc}(r)$.

The numerical integration is performed by a simple Runge-Kutta scheme with
adaptive step-size control. Here especially the calculation of $\ddif \velo/
\ddif \tau$ from Eq.\,(\ref{eq:motion}) is affected by numerical difficulties.
The resolution of the model grid in our non-LTE models (typically 70 depth
points) is much lower than what is needed for the integration of the
hydrodynamic equations. For the extraction of $\Gamma$, $k$, and $a$ it is thus
necessary to employ an interpolation algorithm which provides smooth functions
with well-defined derivatives. We use the monotonic spline-interpolation by
\citet{ste1:90} for this purpose. Further difficulties occur when the critical
point is approached, where Eq.\,(\ref{eq:motion}), which has the form $\ddif
\velo/ \ddif \tau = f / g$, has a singularity with $f \rightarrow 0$ and $g
\rightarrow 0$. Close to this point it is necessary to apply the rule of de
l'Hopital for the calculation of $\ddif \velo/ \ddif \tau$.

In practice, the wind structure is iterated until changes in the mass-loss rate
$\Delta\dot{M}/\dot{M} < 10^{-4}$ are obtained, which is typically the case
after 15--30 iterations between non-LTE radiation transfer and hydrodynamics.
The demands on the numerical accuracy of the non-LTE solution are extremely
high.  Already small deviations from radiative equilibrium can have fatal
consequences, because the critical point reacts very sensitively on changes of
the radiative flux. An accuracy better than 0.01\% is therefore mandatory for
the flux-level.  Because of this, the non-LTE solution alone requires a
computation time of the order of one day on a current personal computer, and a
hydrodynamic model like the one presented in Sect.\,\ref{sec:hydrocal} takes a
pure calculation time of the order of one month.

\section{Self-consistent modeling of a WC\,star wind}
\label{sec:hydrocal}

\begin{figure}
  \parbox[b]{0.49\textwidth}{\includegraphics[width=0.47\textwidth]{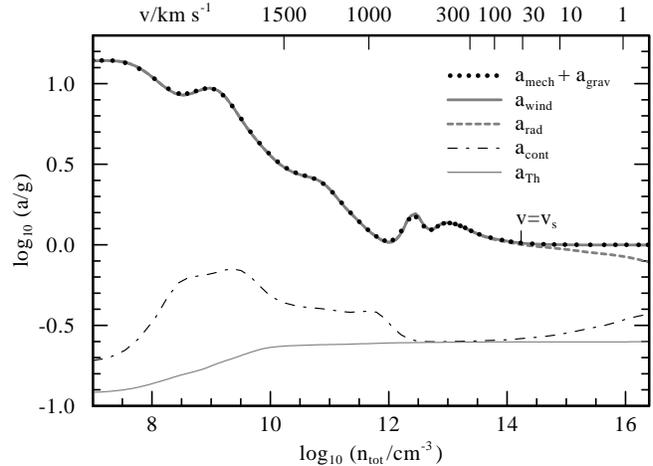}}
\caption []{
  Wind acceleration as in Fig.\,\ref{fig:acc1}, but for the self-consistent {\em
    hydrodynamic} model. The grey line indicates the acceleration
  $a_\mathrm{wind}$ due to radiation + gas pressure as calculated within the
  radiation transport. $a_\mathrm{wind}$ is in precise agreement with the
  mechanical + gravitational acceleration ($a_\mathrm{mech} + a_\mathrm{grav}$,
  black dots) due to the hydrodynamic velocity structure and the stellar mass.
  In addition, the radiative acceleration on free electrons ($a_\mathrm{Th}$),
  on continua ($a_\mathrm{cont}$), and the total radiative acceleration on
  continua and lines ($a_\mathrm{rad}$) are indicated.}
\label{fig:acc3}
\end{figure}

\begin{figure}
  \parbox[b]{0.49\textwidth}{\includegraphics[width=0.47\textwidth]{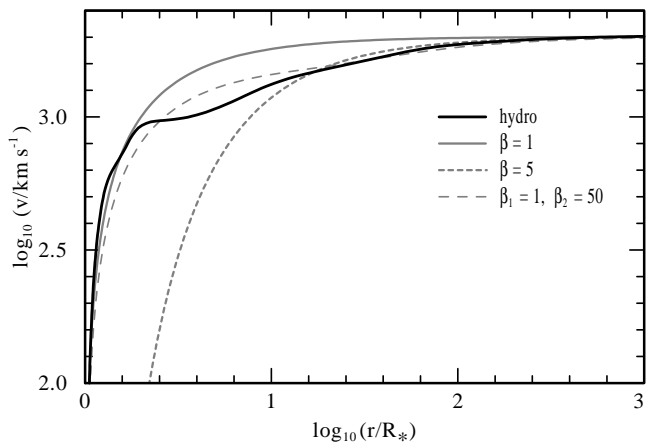}}
\caption []{
  Velocity structure of the {\em hydrodynamic} model (black) compared to
  different $\beta$-type velocity laws and a double-$\beta$ law as proposed by
  \citet{hil1:99}.}
\label{fig:velo}
\end{figure}

\begin{figure*}[] 
  \includegraphics[width=0.99\textwidth]{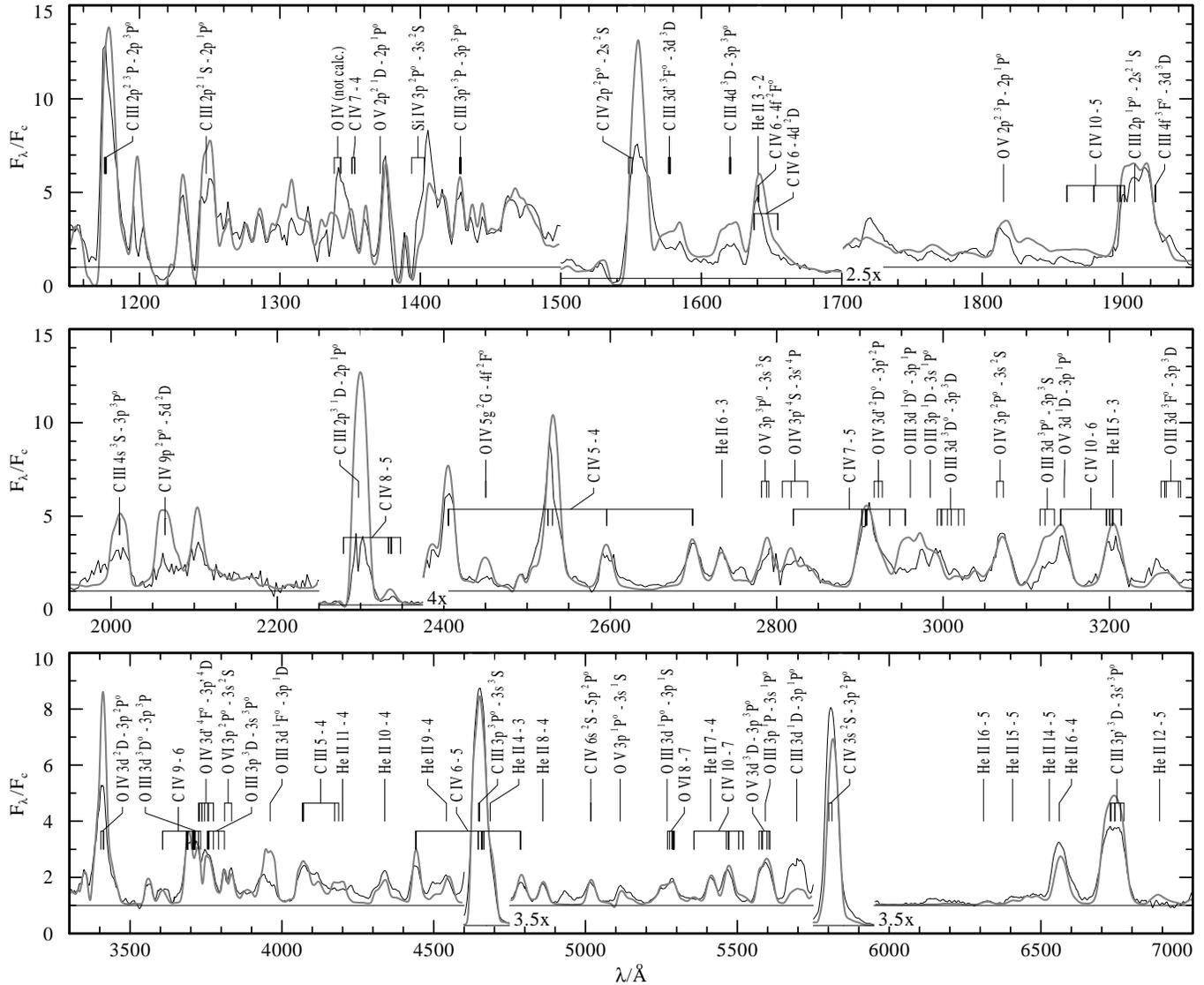}
 \caption{
   Hydrodynamic model: Comparison of the synthetic spectrum (thick line, grey)
   to the observed spectrum of WR\,111 (thin line).  Prominent spectral lines
   are identified.  After correction for interstellar extinction, the observed
   flux and the model flux are both divided by the model continuum.
   Additionally, a correction for interstellar Ly\,$\alpha$ absorption is
   applied. The model parameters are compiled in Table\,\ref{tab:parameters}
   (Hydro). }
\label{fig:spec}
\end{figure*}

\begin{figure*}[tp!] 
  \includegraphics[width=0.97\textwidth]{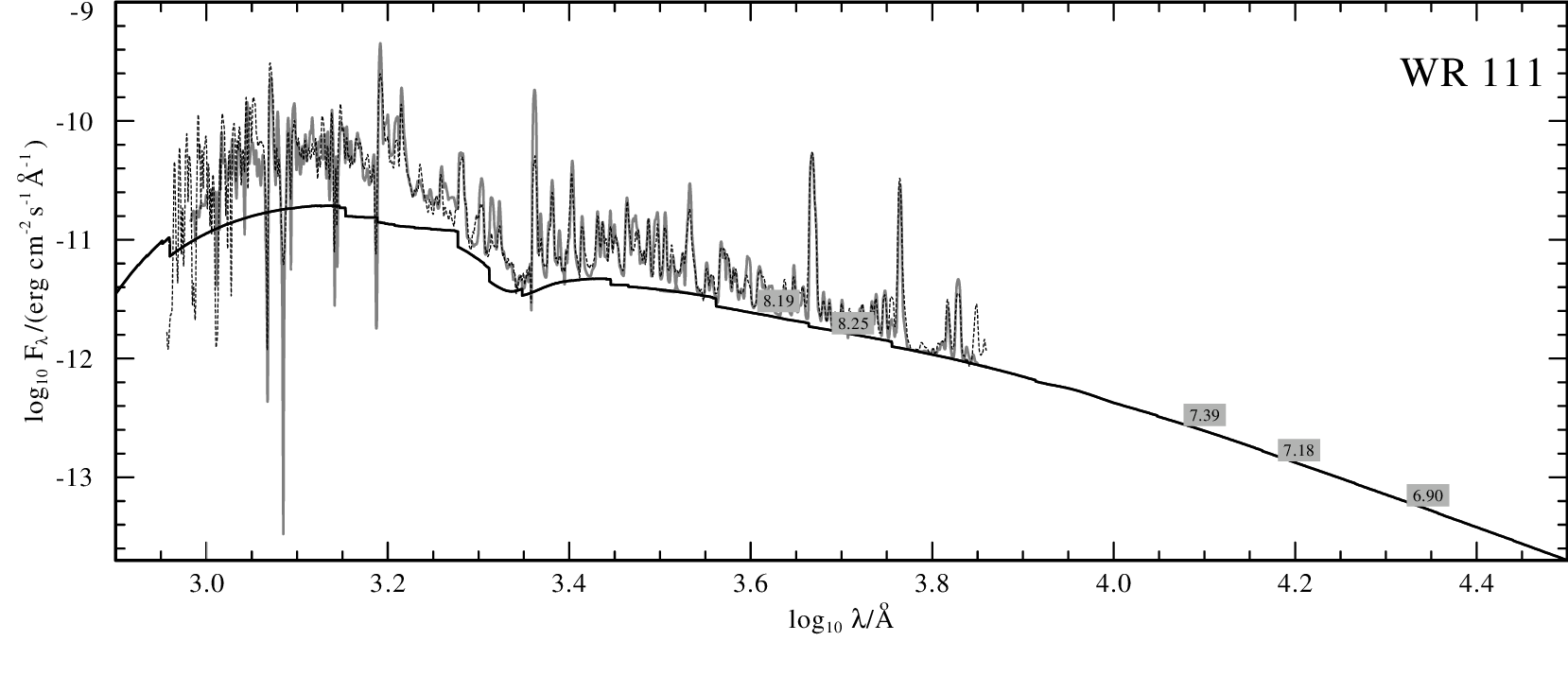}
  \caption{
    Hydrodynamic model: Emergent flux distribution (grey) compared to the
    observed flux of WR\,111 (dashed black line) including optical and infrared
    photometry (labeled blocks).  A distance modulus of 11.0\,mag
    ($d=1.58\,\mathrm{kpc}$) and interstellar reddening (with $E_{B-V} =
    0.25\,{\rm mag}$) are applied to the calculated spectrum (grey) and the pure
    continuum calculation (solid black line).  Moreover, the model spectrum is
    scaled down by 0.18\,dex to match the absolute value of the observed flux.}
  \label{fig:flux}
\end{figure*}

In this section we apply the hydrodynamic treatment described in
Sect.\,\ref{sec:hydro}, and present a self-consistent atmosphere model for an
early-type WC\,star. The `input' parameters for the hydrodynamic model are the
same as for our test model D in Sect.\,\ref{sec:testcal}. Moreover, we assume a
stellar mass of $13.63\,M_\odot$ according to the mass-luminosity relation by
\citet{lan1:89}.  The clumping structure is prescribed in the same way as for
our previous calculations, i.e., $D(\tau_{\rm R})$ is related to the
LTE-continuum optical depth $\tau_{\rm R}$ (see Sect.\,\ref{sec:mpar}) which is
now updated after each iteration.

The converged model nicely reproduces the observed parameters of a typical
WC\,star (see Table\,\ref{tab:parameters}, `Hydro').  The obtained mass-loss
rate is only slightly higher than the value adopted for Model~D, and the
resultant terminal velocity is very close to the canonical value of 2000\,\kms.
In Figs.\,\ref{fig:spec} and \ref{fig:flux} the synthetic spectrum is compared
to the observed spectrum of WR\,111. Although the fit quality would not be
sufficient for a detailed analysis, the model clearly displays the spectral
characteristics of an early-type WC\,star.  Moreover, the successful
reproduction of \OVI\,3811/38 supports our assumptions concerning the radial
dependence of the clumping factor (see previous section), although the electron
scattering wings of \CIV\,5805 indicate that the assumed maximum value of $D=50$
is probably too high.

{ Further discrepancies are mainly due to the modified model parameters
  in comparison to our previous analysis of WR\,111 \citep{gra1:02}. Due to the
  high assumed carbon abundance, many carbon lines appear too strong. Moreover,
  the emergent flux distribution is slightly changed, which results in a smaller
  required correction for interstellar extinction ($E_{B-V}=0.25\,\mathrm{mag}$
  compared to the previous value of $E_{B-V} = 0.325\,\mathrm{mag}$). In the
  present work, the model flux must therefore be scaled down to match the
  observed flux distribution of WR\,111.}
A comparison of the emergent flux distributions is shown in
Fig.\,\ref{fig:flux}, where the model flux has been reduced by 0.18\,dex and a
distance of $d=1.58\,\mathrm{kpc}$ ($M - m = 11.0\,\mathrm{mag}$) is adopted for
WR\,111.

The wind acceleration of our self-consistent model is shown in
Fig.\,\ref{fig:acc3}.  A precise agreement between the mechanical +
gravitational acceleration (black dots) and the acceleration due to radiation
and gas pressure (grey, in the background) is obtained.  Note that the wind
shows two acceleration regions, one in the optically thick part below a velocity
of 900\,\kms, and one in the outer part of the wind. These two zones reflect the
influence of the two well-known opacity bumps, which are responsible for a
variety of phenomena in stellar physics \citep[see][]{rog1:98,nug1:02}. In the
present case, both opacity bumps are located beyond the sonic point and thus
contribute to the acceleration of the wind.

In Fig.\,\,\ref{fig:velo}, the resulting velocity structure is compared to
$\beta$-type velocity laws with $\beta=1$ and $\beta=5$, and a double-$\beta$
law as proposed by \citet{hil1:99} ($\vinf = 2000\,\kms$, $\velo_{\rm ext} =
400\,\kms$, $\beta_1 =1$, and $\beta_2 = 50$). Clearly, none of the $\beta$-laws
is able to describe the overall structure of the hydrodynamic model exactly.
The strong acceleration in the inner part is best described by $\beta=1$ whereas
the outer part tends towards higher $\beta$, in accordance with \citet{lep1:99}
who deduced high values ($\beta > 3$) for the outer wind region from variations
in emission line-profiles. The best match, especially for the asymptotic
behavior at large radii, is indeed provided by \citeauthor{hil1:99}'s
double-$\beta$ law.

\begin{figure}
  \parbox[b]{0.49\textwidth}{\includegraphics[width=0.47\textwidth]{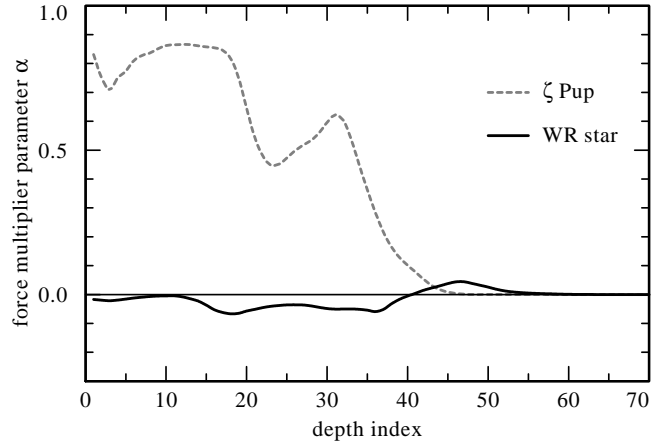}}
\caption []{
  The {\em effective} force multiplier parameter $\alpha$ as calculated within
  the radiation transport {\em vs}. radial depth index (1: outer boundary, 70:
  inner boundary). For the hydrodynamic WR\,model (black) values close to zero
  are obtained, indicating a weak dependence of the radiative force on $\velo'$.
  A model for the O\,star $\zeta$\,Pup \citep{gra2:03} shows values around 0.7
  in the outer part of the wind, as expected from the CAK theory.}
\label{fig:alpha}
\end{figure}

For the parametrization of the line force in the hydrodynamics, the effective
force multiplier parameter $\alpha$ is calculated within our radiation transport
(see Eq.\,\ref{eq:acak}). In the standard picture, $\alpha$ denotes the fraction
of the line force which is due to optically thick lines, i.e., the fraction
which reacts linearly on changes of $\velo'$. As demonstrated in
Fig.\,\ref{fig:alpha}, we obtain values for our WR\,model of $-0.1 < \alpha <
0.1$ throughout the whole atmosphere. This implies that the dependence of the
line force on the velocity gradient is very small and sometimes even shows the
opposite behavior as expected (i.e.\ the line force decreases for increasing
$\velo'$). The displayed O\,star model, in contrast, shows the expected behavior
with $\alpha \approx 0.7$ in the outer wind regions, and $\alpha \rightarrow 0$
at large optical depth.  The abnormal values do presumably {\em not} reflect the
original distribution of line strengths.  Instead, the {\em effective} value of
$\alpha$ is probably influenced by complex radiative processes like line
emission cascades due to recombination in the outer wind regions or extreme line
overlaps.  Hence our results indicate {\em the total breakdown of the CAK theory
  for WR\,winds}, even in the outer wind regions.

Concerning the driving mechanism of the wind, our models are in qualitative
agreement with the ideas of \citet{nug1:02}. The critical point of the equation
of motion (at $\velo_\mathrm{c}=32\,\kms$) is indeed located very close to the
sonic point (at $\velo_\mathrm{s}=28\,\kms$) in the optically thick part of the
envelope, {\em below} the iron opacity peak where the Rosseland mean opacity is
increasing with radius.  The radiative acceleration on the iron opacities
overruns the gravitational attraction close to the sonic radius \citep[see
also][{}\,Sect.\,4.1]{nug1:02}, i.e., the star crosses the Eddington limit at
this point and thereby initiates the mass loss.

The detailed structure of our present model, however, shows significant
differences to the results of \citeauthor{nug1:02} for WR\,111.  This is
basically due to the high value \citeauthor{nug1:02} adopted for the stellar
radius { \citep[corresponding to $T_\star = 81$\,kK, see][]{lam1:02}}.
Because \citeauthor{nug1:02} only analyze the sonic point, and {\em assume} that
the wind is radiatively driven in the outer regions, they obtain solutions for a
relatively wide range of $T_\star$.  Our test calculations, in contrast,
indicate that for `cool' $T_\star$ around $85\,\mathrm{kK}$ the radiative force
is {\em not} sufficient to maintain the WR-type mass-loss beyond the sonic point
(see Model~B in Sect.\,\ref{sec:testcal}).  For our `hot' models with
$T_\star=140\,{\rm kK}$ this situation improves. Consequently, as the
temperature gradient increases proportional to $T_\star$, we obtain higher
temperatures at smaller optical depths.  The sonic point in our hydrodynamic
model is therefore located at a relatively small flux-mean optical depth of
$\tau_{\rm s} = 5.4$ and a high temperature of $T_{\rm s} = 199\,{\rm kK}$,
compared to the models by \citeauthor{nug1:02} where $\tau_{\rm s} = 15$--$25$
and $T_{\rm s} = 140$--$190\,\mathrm{kK}$.

In our hydrodynamic models the mass-loss rate is chiefly determined by the
temperature-dependence of the Rosseland mean opacity.  An increase of the wind
density heats up the deeper layers and pushes the opacity bump, and therefore
also the sonic point, towards lower densities. Consequently, the mass-loss rate
is decreased, i.e.\ the wind is stabilized.  This regulation mechanism is most
effective in parameter ranges where the ionization structure is sensitive to
changes in the mass-loss.  This is probably the reason why the WR\,models in
spectral analyses tend to lie in regions where the ionization is just changing
(i.e.\ the models are most difficult to calculate).

Another important question is how our results fit to the predicted properties of
stellar structure models \citep[e.g.\,{}][]{lan1:89,sch1:96}. An inspection of
Figs.\,4 and 5 in the work of \citet{sch1:96} shows that our sonic-point
temperature of $T_{\rm s} = 199\,{\rm kK}$ is already reached relatively deep
inside the star at around $0.8\,R_\star$. Moreover, this point is located within
the outer convection zone of the star, which is also caused by the iron opacity
bump in these models.  This implies that the convection zone in stellar
structure calculations may possibly be identified with the acceleration zone of
the WR\,wind, which then would be located at a rather small radius. However, for
exact predictions stellar structure models are needed that take realistic wind
models as an outer boundary condition, and allow for convection under dynamical
conditions.

\section{Conclusions} 
\label{sec:conclusions}

In the present work we have constructed the first hydrodynamically consistent
model for a WR-type (WC\,5) stellar wind, driven by radiation pressure.  For
this purpose we have combined our non-LTE atmosphere models with the
hydrodynamic equations. On this way we obtain the first models for radiatively
driven winds that incorporate a full non-LTE radiation transfer in the CMF,
i.e., they allow for a realistic treatment of all important radiative effects
throughout the whole atmosphere, including large optical depths.  Moreover, the
models account for the important effects of clumping.

By inclusion of the Fe M-shell ions (\FeIX--\FeXVI) in our models, the radiative
acceleration in deep atmospheric layers is considerably enhanced.  We show that,
for suitable stellar parameters, the radiative force on these opacities is able
to drive an optically thick wind. In particular, very high stellar temperatures
($T_\star\approx 140\,\mathrm{kK}$) are needed to maintain the wind acceleration
above the sonic point. In accordance with \citet{nug1:02}, the sonic point in
our hydrodynamic model is located at large optical depth {\em below} the `Hot
Iron Bump' at a very high temperature of $T_{\rm s} = 199\,{\rm kK}$.  The
resultant wind structure shows two acceleration regions.  A steep velocity
increase in deep atmospheric layers due to the Fe M-shell ions, and a shallow
increase in the outer part of the wind due to iron-group ions of lower
excitation in combination with C and O. The latter is in accordance with the
shallow velocity laws deduced by \citet{lep1:99} from line-profile variations.

Despite of the fact that the wind is optically thick at the sonic point, we find
strong deviations from the standard CAK theory also in the outer wind regions.
Here especially the reaction of the radiative force on changes of $\velo'$ shows
an abnormal behavior, which is presumably due to complex radiative processes
like line emission cascades after recombination, or extreme line overlaps.
Moreover, we find a strong dependence of the radiative force on wind clumping,
an effect which is neglected in previous models for radiatively driven winds.

The emergent flux, as calculated from our hydrodynamic model, clearly shows the
spectral characteristics of an early-type WC\,star, i.e., the wind density and
the terminal wind velocity, which are both obtained self-consistently from the
hydrodynamics, are of the correct order of magnitude.  Moreover, we are able to
reproduce the observed narrow {\OVI} emissions in the optical spectrum of
WC\,stars. These features are excited by a combination of heating due to the Fe
M-shell ions and the high stellar temperature, which is necessary for our wind
modeling.  The {\OVI} lines also react very sensitively on the detailed clumping
structure, i.e., they help to trace deep atmospheric layers where the clumping
of WC\,star winds sets in.

When our results are compared to stellar structure calculations, the question
about the location of the `surface' of a WR\,star gains more and more in
importance. Stellar structure models for WR\,stars show an outer convection
zone, which largely determines the extension of the stellar envelope. The
convection however, is caused by the same increase of the mean opacity that
causes the driving of the WR\,wind in our models. A temperature around 200\,kK,
corresponding to our sonic-point temperature, is only reached deep inside the
stellar envelope at the {\em bottom} of the outer convection zone.  For a full
understanding of the structure and the mass-loss of WR\,stars models are
therefore needed, that combine stellar structure and atmosphere calculations and
allow for convection under dynamic conditions.

\begin{acknowledgements} 
  This work was partly supported by the Deutsche Agentur f\"ur
  Raumfahrtangelegenheiten under grant \mbox{DARA 50\,OR\,0008} and partly by
  the Swiss National Science Foundation.
\end{acknowledgements}

\bibliographystyle{aa} \bibliography{aamnem99,astro}
 
\end{document}